\documentclass[twocolumn,showpacs,epsfig,preprintnumbers,amsmath,amssymb]{revtex4}

\usepackage{units,xspace}
\usepackage{graphicx}
\usepackage{dcolumn}
\usepackage{bm}
\def \be{\begin{equation}}
\def \ee{\end{equation}}
\def \bea{\begin{eqnarray}}
\def \eea{\end{eqnarray}}
\def \nn{\nonumber}
\def \ben{\begin{enumerate}}
\def \een{\end{enumerate}}

\newcommand{\vecr}{{\bf r}}
\newcommand{\vecv}{{\bf v}}
\newcommand{\vecF}{{\bf F}}

\newcommand{\textf}{\text{F}}
\newcommand{\vecH}{{\bf H}}

\begin{document}

\title{Collective excitations of hydrodynamically coupled driven colloidal particles}
\author{Harel Nagar and Yael Roichman}
\affiliation{Raymond \& Beverly Sackler School of Chemistry, Tel Aviv
  University, Tel Aviv 6997801, Israel}

\date{\today}

\begin{abstract}

Two colloidal particles, driven around an optical vortex trap, have been recently shown to pair due to an interplay between hydrodynamic interactions and the curved path they are forced to follow. We demonstrate here, that this pairing interaction can be tuned experimentally, and study its effect on the collective excitations of many particles driven around such an optical trap. We find that even though the system is overdamped, hydrodynamic interactions due to driving give rise to non-decaying excitations with characteristic dispersion relations. The collective excitations of the colloidal ring reflect fluctuations of particle pairs rather than those of single particles.
\end{abstract}


\maketitle
\section{Introduction}

Driven dissipative systems made of colloidal particles suspended in liquid afford a way to study  many-body non-equilibrium dynamics at the single particle level \cite{beatus06,roichman07,lutz06}. For example, using a 1D crystal of oil droplets driven along a microfluidic channel, non-equilibrium hydrodynamically-induces phonons were recently studied  \cite{beatus07,beatus06,beatus12}. Surprisingly, the crystal vibrations, in that system, did not decay due to thermal fluctuations, as could be expected at such low Reynolds numbers (Re$\sim10^{-4}$), but continued to propagate acoustically at a sound velocity of $C_s\approx250 \mu$m/s~\cite{beatus06}. The dispersion relations of this 1D crystal were different from those of a harmonic crystal, they exhibited symmetry breaking due to flow directionality. Here we study the collective excitations of a different driven dissipative system. In our system particles are driven along a ring-like optical trap, which means that they are confined to a quasi one dimensional track, but interact hydrodynamically in 3D rather than in 2D as in the previously studied system \cite{beatus06}. Moreover, particles in our system tend to pair due to a hydrodynamically induced pseudo-attraction \cite{sokolov11}. We show here, that this pairing interaction greatly affects the particles' collective excitations.

 \begin{figure}[h!]
    \centering
    \includegraphics[width=0.95\columnwidth]{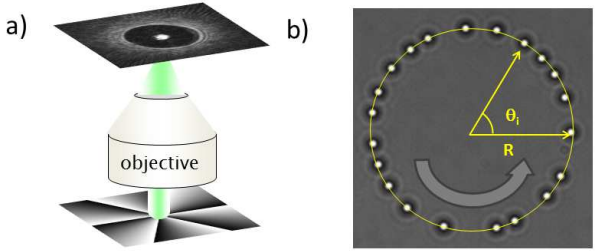}
    \caption{An optical vortex laser trap. a) a phase pattern is imprinted on a laser beam in a plane conjugated to the sample plane. A microscope objective is then used to focus the beam tightly, forming a ring-like optical trap. b) 1.1 $\mu$m polystyrene beads trapped in the optical vortex rotate at approximately constant angular velocity.}
    \label{fig:system}
 \end{figure}
 
We use holographic optical tweezers (HOTs) \cite{molloy02,grier03} to confine suspended colloidal particles and drive them along a 1D circular optical trap. Thus particles never leave the field of view, and can be followed for long periods of time (i.e. an hour). One type of such a circular trap is created with HOTs by projecting  a helical beam of light (optical vortex), carrying angular momentum \cite{curtis03,padgett00}, into the sample plain (Fig \ref{fig:system}). When particles are trapped in an optical vortex beam, this angular momentum is transfered to them, causing them to rotate around the ring-shaped optical trap \cite{ladavac04,sokolov11,sassa12}. As a result, the trapped colloidal particles constitute a driven dissipative system in which the particles are driven at constant angular velocity and are coupled to a thermal bath (the suspending media). Many properties of this model system have bean calculated \cite{reichert04,ladavac05} and measured \cite{reichert04,roichman07,sokolov11, sassa12} in the past two decades. On a broader context, a single particle driven along a ring trap constitutes a model for the motion of cilia \cite{Reichert2005,niedermayer08}, and has been used to study synchronization among cilia hairs in arrays of ring traps (e.g. \cite{Reichert2005,niedermayer08,koumakis13,kotar13}). The same mechanism that is responsible for particle pairing of two particles rotating in a single ring \cite{sokolov11}, causes synchronization between single particles rotating in two adjacent rings \cite{niedermayer08}. In both cases the effect depends on the ratio between tangential driving and radial confinement along the ring.   
 
The equations of motion for $N$ suspended colloidal particles subjected each to a force, $\vecF_i$, and interacting hydrodynamically, are given by \cite{polin06}:
\be
\vecv_i=\sum\limits_{j=1}^{N}\vecH_{ij}\vecF_j.
\label{Eq:vel}
\ee
where $\vecv_i$ is the velocity of particle $i$. It is common practice to assume pairwise additivity of the hydrodynamic interactions \cite{polin06,Cicuta10,reichert04} in the stokeslet approximation to derive $\vecH_{ij}$; 
\be
\vecH_{ij}\approx\frac{{\bf I}}{\gamma}\delta_{ij}+\frac{3}{4\gamma}\frac{a}{r_{ij}} \left({\bf I}+\vecr_{ij}\otimes\vecr_{ij} \right)(1-\delta_{ij})
\label{Eq:stokes}
\ee  
where $\gamma=6\pi\eta a$ is the stokes drag on the sphere, and $\eta$ is the dynamic viscosity of the fluid.
In an optical vortex the optical forces acting on a typical particle are a constant tangential driving force $\textf_\theta\hat{\theta}$, and a radial confining force $\textf_r=k_r(r_i-R)\hat{r}$, where $k_r$ is the trap stiffness, $r_i$ is the radial position of particle $i$, and $R$ is the radius of the optical vortex. These equations of motion were linearized, at zero temperature, around a configuration of equidistant arrays of trapped particles in \cite{reichert04}. The angular velocity of the particles, $\dot{\theta}_i= \Omega+\delta\dot{\theta}_i$, was separated into a constant value, $\Omega$ and a time varying component, $\delta\dot{\theta}_i$. Assuming the radial confinement to be infinitely strong, $\Omega$ and $\dot{\delta\theta_i}$ are given by:  
\bea
\Omega &=& \frac{\textf_\theta}{6 \pi \eta a R}\left(1+\frac{3a}{8R}\sum\limits_{i\neq j}\frac{1+3\cos(\Delta\theta_{ij})}{\sqrt{2-2\cos(\Delta\theta_{ij})}}\right)  \\
\dot{\delta\theta_i} &=&  -\frac{\textf_\theta}{16 \pi \eta R^2}  \sum\limits_{i\neq j}\frac{(7-3\cos(\Delta\theta_{ij}))\sin(\Delta\theta_{ij})}{(2-2\cos(\Delta\theta_{ij}))^{3/2}}\delta\theta_j \nn
\label{Eq:motion}
\eea
where $\Delta\theta_{ij}=\theta_j-\theta_i$, $\eta$ the fluid viscosity, and $a$ the radius of the colloidal particles.
The solution of these equations reveals several oscillatory modes of motion which the system can adopt \cite{reichert04}. In experiments, trapped colloidal particles seem to alternate between the different modes \cite{sassa12,roichman07}. Note that these modes are inherently different from the hydrodynamic modes that exist in equilibrated arrays of trapped colloids \cite{polin06,Cicuta10} as they have real eigenvalues \cite{reichert04,sassa12}.

It was recently shown \cite{sokolov11} that when only two particles are trapped in an optical vortex, they pair. This pairing attraction is a result of symmetry breaking induced by the curved path the particles are forced to take. Consequently, the average angular velocity of particles in an $N$ particle vortex can be predicted from the zero mode motion of the aforementioned equations of motion \cite{reichert04}, by treating each particle pair as an effective particle with an effective mobility \cite{sokolov11}. One important aspect of the pairing interaction is that it is a pure non-equilibrium effect, it arises only through  driving. This means that the effects of driving are observable even in the co-rotating frame, i.e. the formation of particle pairs.
Particle pairing can be overcome in two ways: by increasing temperature, or by increasing the ratio between the radial confinement force and the tangential driving force, $\kappa=\textf_{r}/\textf_{\theta}$ \cite{sokolov11}.

In this paper we study the collective excitations of particles trapped in an optical vortex as a function of the pairing interaction strength. We start by demonstrating, for the first time, that $\kappa$ can be controlled by tuning laser powers. We then correlated between the trapping laser power and $\kappa$, in a one-particle vortex, and find a range in which $\kappa$ changes linearly with laser power. We proceed to verify the assumption that $\kappa$ controls pairing by measuring the average angular separation between two particles trapped in an optical vortex as a function of $\kappa$ measured from single particle experiments. Finally, we turn our attention to the collective vibrational modes of many particles trapped in an optical vortex. We start by calculating theoretically and numerically the dispersion relations of particles in a vortex as a function of the driving force and particle number. We then compare our calculations to the experimentally measured dispersion relations and show that they are non-decaying, similar to what was found for a 1D crystal of oil droplets \cite{beatus07}, but are better described as fluctuations of particle pairs rather than as fluctuations of single, separated, beads. 

\section{Methods}

{\bf Experiments}- Colloidal suspensions of polystyrene beads,  1.1 $\mu m$
in diameter (Invitrogen lots \#742530), were place in between a glass slide and a cover-slip to form an approximately 40~$\mu m$ thick sample. During experiments the suspended colloidal particles were subjected to an optical vortex trap generated by holographic optical tweezers \cite{dufresne01,curtis03,sokolov11}. In our setup a Gaussian laser beam (Coherent, Verdi $\lambda=532$~nm) is transformed into a helical beam carrying a topological charge of $\ell=100$ using a spatial light modulator (Hamamtsu, X10468-04). The beam is then inserted into an inverted microscope (Olympus, IX71) and focused through the microscope objective (100x oil immersion NA=1.4) forming a ring-like optical trap with a diameter of $9.1\pm0.2$~$\mu m$ and diffraction limited thickness. We performed three sets of experiments trapping either 1,2 or 21 particles, driving them at laser powers of 0.5-4 W (at the output of the laser), and imaging them at either 10 fps (Hamamtsu,Orca flash 4), or at 70.4 fps (Gazzele, Point gray). Particle tracking was done using conventional video microscopy \cite{crocker96}. During each experiment particles were allowed to reach a steady state before measurement commenced.
 
{\bf Simulations}- We use the Stokesian Dynamics protocol \cite{Brady88} to simulate particles driven in an optical vortex due to its tested ability to correctly account for many particle hydrodynamic interactions, and its incorporation of thermal forces. Extending Eq.~\ref{Eq:vel} to include thermal forces and replacing the Stokeslet approximation (Eq.~\ref{Eq:stokes}) with the Rotne--Prager operator \cite{RotnePrager}, we have:
 \begin{equation}
\vecv_i =
\sum_{j=1}^{N}\frac{{\bf D}_{ij}}{k_BT}\cdot\vecF_j +
{\bf \zeta}_i,
\label{eq:langevin}
\end{equation}
where ${\bf D}_{ij}=k_BT {\bf H_{ij}}$ the diffusion tensor, $k_B$ Boltzmann's constant, $T$ the absolute temperature, and ${\zeta}_i^{\alpha}$ represents the stochastic thermal force:
\begin{equation}
\langle{\zeta}_i^{\alpha}(t,{\bf r}_i)
{\bf\zeta}_j^{\beta}(t',{\bf r}_j)\rangle
=2{D}_{ij}^{\alpha\beta}\delta(t-t').
\label{eq:zeta}
\end{equation}
We obtain 
$\{\zeta_i^{\alpha}\}$ by employing the Cholesky decomposition on the diffusion matrix: ${\bf D=AA^T}$, with ${\bf A}$ the lower triangular matrix, and $\{R_i^{\alpha}\}$ the basis set of random Gaussian variables $\langle R_{i}^{\alpha}(t) R_{j}^{\beta}(t')\rangle =
 \delta(t-t')\delta_{ij}\delta_{\alpha\beta} $,
\begin{equation}
 \zeta_{i}^{\alpha} = 2A_{ij}^{\alpha\beta}R_j^{\beta}.
 \end{equation}
The discretized equations of motion then read, 
\begin{equation}
{\bf r}_i(t+\Delta t)={\bf r}_i(t)
+ {\Delta t}\sum_{j=1}^{N}\frac{{\bf D}_{ij}}{k_BT}\cdot\vecF_j
+\sqrt{\Delta t}\sum_{j=1}^{N}
{\boldsymbol A}_{ij}\cdot{\boldsymbol R}_j.
\label{eq:stoke_dyn}
\end{equation}
with the Rotne--Prager tensor \cite{RotnePrager} given by:
\begin{equation}
 \frac{{D}_{ij}^{\alpha\beta}}{D_0}=\left\{
 \begin{array}{lr}
\delta_{\alpha\beta} & \text{if } i=j\\ 
\frac{3}{4}\frac{a}{r_{ij}}
\Bigg(\delta_{\alpha\beta}+\frac{x_{ij}^{\alpha}x_{ij}^{\beta}}{r_{ij}^2}\Bigg)+ & \text{if } i\ne j\\ 
\text{       }+\frac{1}{2}\frac{a^3}{r_{ij}^3}
\Bigg(\delta_{\alpha\beta}
-3\frac{x_{ij}^{\alpha}x_{ij}^{\beta}}{r_{ij}^2}\Bigg)
 \end{array}
\right.
\end{equation}
with $D_0=k_BT/\gamma$.

In the simulations particles are allowed to move in two dimensions confined to a ring of radius $R$ with a harmonic restoring force $\textf_r=k_r \Delta \vecr \hat{r}$, and a driving force $\textf_r=k_\theta R \hat{\theta}$. A Weeks-Chandler-Anderson repulsive pair potential \cite{Weeks71} is used to represents the hard core repulsion between colloidal particles. Simulation parameters were chosen to match experiments in terms of particle size, vortex radius, driving and confining forces and temperature. The time interval was equivalent to $dt = 0.001 s$.   
There are two main differences between the experimental system and the simulated system. Experiments are conducted near a glass wall whereas simulations assume no near boundary. As a result the hydrodynamic interactions in experiments should be weaker. In addition, in simulations the optical potential landscape is perfectly smooth, unlike the situation in experiments.
  \section{Particle pairing}

\begin{figure}[h!]
       \centering
       \includegraphics[width=0.95\columnwidth]{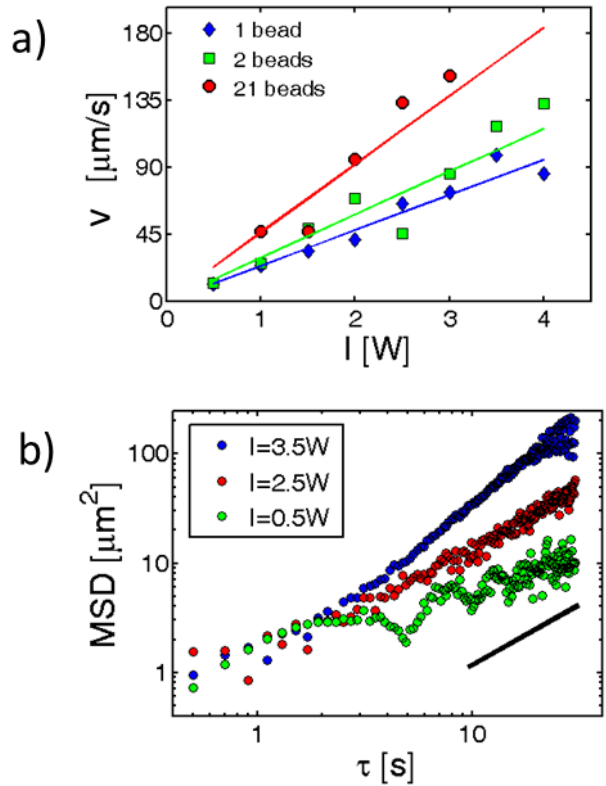}
       \caption{a) Mean angular velocity of particles in an optical vortex as a function of driving power for systems with 1,2, and 21 trapped beads. b) The mean square displacement in the tangential direction of a single particle in the co-rotating frame at different driving powers.}
       \label{fig:system2}
    \end{figure}

 A colloidal particle driven along an optical vortex experiences, ideally, two types of optical forces: an intensity gradient force confining it radially $\textf_r\propto\nabla I\hat{r}$, and a phase gradient force, due to radiation pressure, driving it tangentially  $\textf_\theta\propto I\nabla\varphi\hat{\theta}$ \cite{reichert04,roichman08}. We measure these forces by tracking a single particle rotating in an optical vortex (Fig.~\ref{fig:system2}a).
 
 The tangential driving force can be extracted from the average angular velocity of the bead $\Omega$, since $\Omega=\textf_{\theta}/\gamma R$. 
 The radial confining force $\textf_r$, can be extracted from the radial position distribution of the bead, using $\textf_r=k_r w/2$, where $k_r=K_B T/\langle(\Delta r)^2\rangle$ is the stiffness of the trap in the radial direction, $w\sim 0.25 \mu$m is its width, and  $\langle(\Delta r)^2\rangle$ is the variance in the bead's radial position.

 \begin{figure}[h!]
    \centering
    \includegraphics[width=0.95\columnwidth]{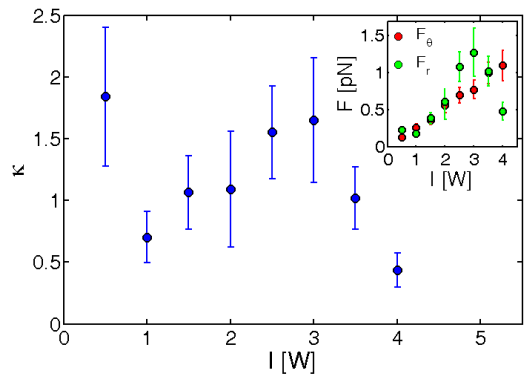}
    \caption{The ratio between radial trapping and tangential driving $\kappa$, as a function of laser power. Inset shows the radial $\textf_r$ and tangential $\textf_\theta$ forces.}
    \label{fig:kappa}
 \end{figure}

The optical potential landscape of an optical vortex trap is not completely uniform \cite{sokolov11}, due to imperfections in the optical train. As a result, the particles move over a rugged landscape, similar to the motion over a periodic potential with a small amount of randomness (for example see theoretical treatment in \cite{Khoury11}).
The roughness of the optical potential is characterized indirectly, by studying the tangential diffusion of a single particle (Fig.~\ref{fig:system2}b) at the co-rotating frame (the frame of reference rotating at the mean angular velocity of the particle). We observe that at low laser powers the particle diffusion becomes sub-diffusive ($\langle(\Delta r)^2\rangle \sim t^\alpha$ with $0<\alpha<1$), while at high laser powers it is super-diffusive ($\langle(\Delta r)^2\rangle \sim t^\alpha$ with $1<\alpha<2$). We attribute these changes to the effect of the rough optical potential. At intermediate laser powers the particle diffuses normally, which implies that the landscape effect is minimal. We therefore perform our experiments in this regime.      

In Fig.~\ref{fig:kappa} the forces $\textf_r$, $\textf_\theta$, and their ratio, $\kappa$, are shown as a function of laser power. Here the same dynamical regimes are observed, at low driving force $\kappa$ is large and widely distributed. At high driving force $\kappa$ decreases with laser intensity and has a narrow distribution. At intermediate driving forces ($1W\le I \le 3W$),  where the particle diffusion is normal (see Fig.~\ref{fig:system2}b), and we perform our experiments, $\kappa$ increases linearly with laser intensity.   

 \begin{figure}[h!]
   \centering
   \includegraphics[width=0.99\columnwidth]{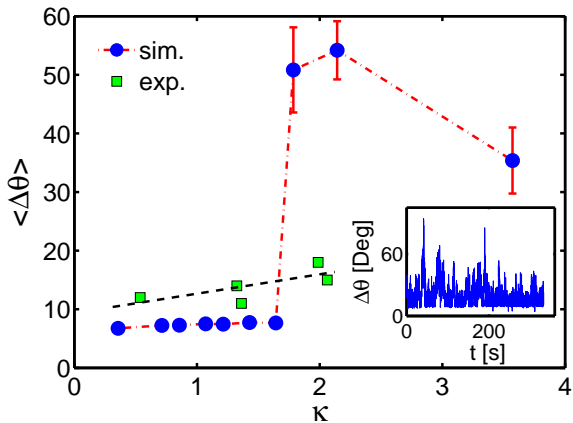}
   \caption{Angular separation of two beads trapped in an optical vortex as a function of $\kappa$, the ratio between the radial confinement force and the tangential driving force for simulations (blue circles), and experiments (green squares). In experiment laser powers vary between $1W \le I \le 3W$.  The inset shows a typical time series from experiment of the angular separation between two particles (I=2~W, $\kappa=1.55$), the pairing angle was averaged over the paired state, e.g. in the inset, of angles lower than $30^o$.}
   \label{fig:dtheta}
 \end{figure}

 When a second particle is added to the optical vortex trap, the system relaxes to a new steady state in which the two particles come close and their angular separation $\Delta\theta$ fluctuates around some mean value. The inset of Fig.~\ref{fig:dtheta} shows a typical time series of $\Delta\theta$ exhibiting significant fluctuations during experiments. From simulations and previous work  \cite{roichman07,sokolov11} we deduce that our experiments are performed at an intermediate temperature regime in which the system translates between paired and unpaired states continuously. The angular separation between the particles has a main peak with secondary lobes. We define a paired state as a state in which the angular separation lies in the main peak, and the pairing angle as the average angular separation between the particles at such a paired state. 
 At a low laser power of $I=0.5$W no distinct particle pairing is observed, i.e. the inter-particle distance varies significantly with time. Increasing laser power to $1W\le I \le 3W$, corresponding to $0.54\le \kappa\le 2.06$, particles become paired (see Fig.~\ref{fig:dtheta}). The average separation angle,   $\langle\Delta\theta\rangle$, depends linearly on $\kappa$, regardless of laser power (compare to Fig.~\ref{fig:kappa}), and ranges from $\langle\Delta\theta\rangle=10^o$ to $\langle\Delta\theta\rangle=18^o$. This corresponds to distances of 1.4 to 2.6 particle diameters, meaning that although particles come close to touching during experiments, occasionally, they are further apart at their preferred angular separation. In simulations (Fig.~\ref{fig:dtheta}) pairing angle decreases with increasing $\kappa$ as well, however, unlike the experimental findings, the pairing angle show particles almost touching. In addition, above a certain $\kappa$, the particles no longer pair.  This crossover is not experimentally accessible.
 
\section{Collective dynamics}

\begin{figure}[h!]
   \centering
   \includegraphics[width=0.95\columnwidth]{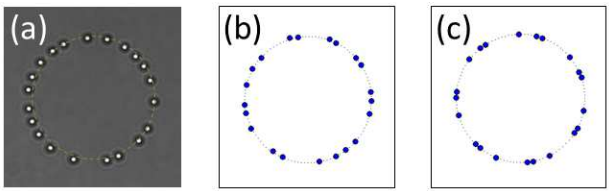}
   \caption{21 particles trapped in an optical vortex. a) Microscope image of polystyrene spheres, 1.1 $\mu m$ in diameter, trapped in an optical vortex of topological charge $\ell=100$, at $\kappa=0.54$. b) Simulation results for similar conditions and room temperature, and c) simulation results at zero temperature.}
   \label{fig:particles}
 \end{figure}

We now turn our attention to the collective excitation of particles on an optical vortex. In our system, a vortex of topological charge $\ell=100$ can hold as many as 21 particles for a long period of time (several hours). This means that the 1D volume fraction of particles is $\phi=21a/2\pi R=0.4$. A larger number of particles can be trapped in the ring, but with time particles start leaving the vortex, rotating alongside other particles, and reentering the trap occasionally \cite{supp}. We note here that odd numbers of particles in a vortex beam, at high density, seem more stable than even numbers at similar densities, in experiments.
Taking into account our results from the one- and two-particle experiments, we choose to work at intermediate laser powers ($1W \le I \le 3W$). In this range an increase in laser power corresponds to a decrease in pairing attraction. The mean velocity of the particles increases with particle number \cite{sokolov11} and with driving power (see Fig.~\ref{fig:system2}a),
ranging from $v=12\mu$m/s to $v=150\mu$m/s for our 21 particle vortex. 
In Fig.~\ref{fig:particles} a snapshot of 21 particles rotating in an optical vortex at $\kappa=0.54$ is compared with snapshots from simulation at room temperature  and zero  temperature. The colloidal configuration at low temperature shows a periodic structure, that is part of a limit cycle motion (see for example supplementary movie \cite{supp2}) including many particle pairs. At room temperature the particles' configuration (Fig.~\ref{fig:particles}b) and dynamics  is not as ordered (see supplementary movie \cite{supp3}). From these images it is not clear if, and to what extent, the hydrodynamic induced pairing interaction affects the collective particle excitations, especially at room temperature.     

\begin{figure}[h!]
   \centering
   \includegraphics[width=0.95\columnwidth]{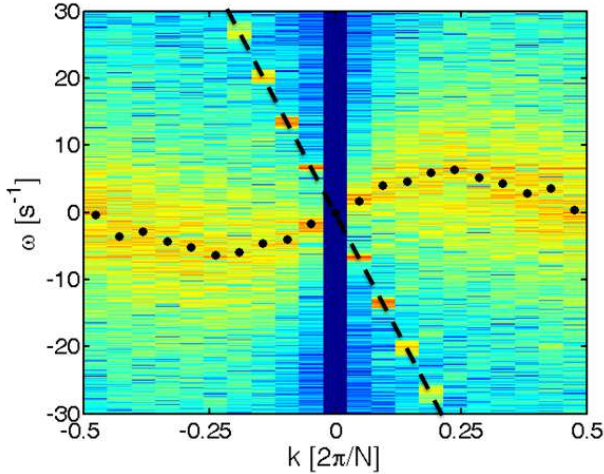}
   \caption{The dispersion relations,
   $\omega(k)$, for longitudinal waves at $\kappa=1.33$. Two branches are seen, a sine-like curve due to hydrodynamic interactions between colloidal particles (the peak position of which is present by black dots), and a straight line due to the response of particles to imperfections in the optical vortex (dashed line).}
   \label{fig:dispersion}
 \end{figure}

In order to address this question we analyze the fluctuations in the angular position of the particles in the co-rotating frame in terms of their dispersion relations, $\omega(k)$, following the analysis of \cite{beatus06}. We start by defining new coordinates, the angular separation between adjacent particles,
\[ \Delta\theta_i = \left\{ 
  \begin{array}{l l}
   \theta_1-\theta_{N} & \quad \text{if $i=1$}\\
    \theta_i-\theta_{i-1} & \quad \text{otherwise}
  \end{array} \right.\]
We apply a discrete Fourier transform both on angular separation and time and obtain the power spectrum of particle fluctuations. The dispersion relations are extracted from the peaks of the power spectrum. In Fig.~\ref{fig:dispersion} the experimentally measured dispersion relation of the angular fluctuations of 21 particles driven at $\kappa=1.33$ is presented. Two branches can be observed; a linear relation due to particles responding to intensity variation along the optical vortex (arising from imperfections in the optical train), and a tilted sine-like curve arising from inter-particle hydrodynamic interactions. Similar to the observations for oil droplets in a microfluidic channel \cite{beatus06}, we observe symmetry breaking of the dispersion relations due to driving, acoustic wave behavior near $k=0$ propagating at sound velocities of $180\mu$m/s to $810\mu$m/s, and group velocity sign change within the Brillouin zone.

\begin{figure}[h!]
      \centering
      \includegraphics[width=0.95\columnwidth]{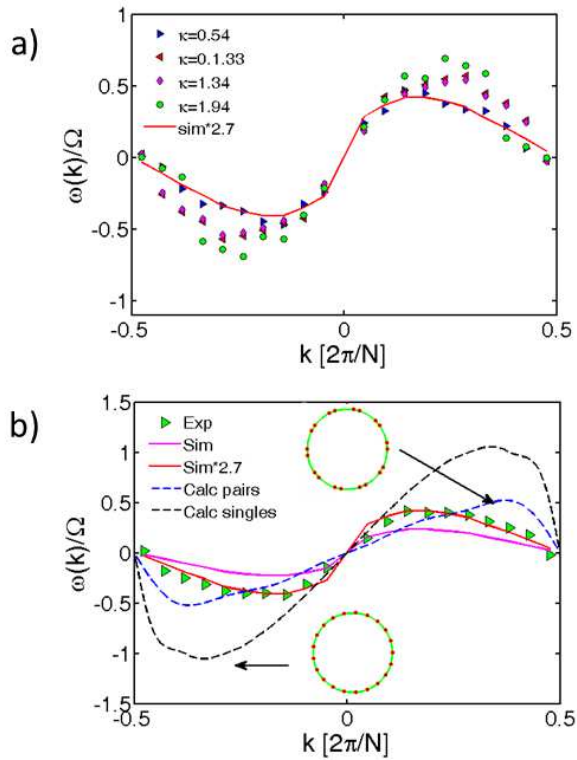}
      \caption{ a) Angular dispersion relation of particles trapped in an optical vortex driven at increasingly high laser powers. As $\kappa$ increases and pairing decreases the experimental curves depart from the simulated curve and approach the theoretical curve for single particles. b) Comparison of experiment, theory and simulation. While the shape of the dispersion relation on the simulated system is identical to that obtained from experimental data, the amplitude is 2.7 times higher in experiment (with $\kappa=0.54$) than in simulation.}
      \label{fig:dispersion_compare}
    \end{figure}

Figure \ref{fig:dispersion_compare}a shows the measured dispersion relation, normalized by the average angular velocity $\Omega$, of the 21 particles driven in the ring at different $\kappa$. The dispersion relation of a simulated system are presented as well, multiplied by a factor of 2.7 to fit the experimental data. We find that in experiments the dispersion relation shape changes significantly as $\kappa$ increases. A much weaker dependence on $\kappa$ was observed in simulation. There are two possible effects that can explain the discrepancy between experiment and simulation: the weaker hydrodynamic interaction in experiments due to the presence of a nearby glass wall, and the experimental rugged optical landscape over which the particles are driven.    

In order to obtain more intuitive understanding of the collective excitations in our system, we calculate theoretically the dispersion relation of particles in an optical vortex. We assume that the equilibrium configuration of particles is a uniform distribution around the ring, and use Eq.~(1) to describe their motion.
Substituting
\be
\delta\theta_m=e^{i(mk\frac{2\pi}{N}-\omega t)}
\ee
into Eq.~(1) and solving for $\omega$, we have
\be
\frac{\omega(k)}{\omega_0} = 2\sum\limits_{m=1}^{\lfloor\frac{N}{2}\rfloor}\frac{7-3\cos(\frac{2\pi}{N}m )}{(2-2\cos(\frac{2\pi}{N}m))^{3/2}}\sin(\frac{2\pi}{N}m)\sin(\frac{2\pi}{N}mk)
\label{Eq:dispersion}
\ee
where $\omega_0=\textf_\theta/16 \pi \eta R^2$, and N is the number of particles trapped on the ring. At low temperatures and low $\kappa$ we expect particles to form stable uniformly distributed pairs \cite{sokolov11}. Assuming that particle pairs are point objects with double the force acting on them we can use the same equation (Eq.~(\ref{Eq:dispersion})) to model their dispersion relations provided we multiple $\textf_\theta$ by 2 and divide the number of particles by 2.
Clearly, neither configuration describes our experiments well, as our particles (and particle pairs) are not point objects, radial confinement is not infinitely strong, and particles' positions fluctuate significantly (see Fig.~\ref{fig:particles}a and movies \cite{supp2,supp3}). Surprisingly, this crude estimation affords insight into the collective excitations observed. 
A theoretical curve for 21 particles arranged uniformly around the vortex (dashed black line) and a curve for 10 particle pairs arranged uniformly around the vortex driven at twice the force (solid blue line) are compared in Fig.~\ref{fig:dispersion_compare}b to experiment and simulation. The theoretical calculations are normalized by the calculated average angular velocity of particles in the vortex. We observe, from Eq.~(\ref{Eq:dispersion}), that the dispersion relation amplitude depends on the force acting on each bead, and its shape on the configuration of the beads. Comparing between experiment and theory, we find that particle pairing affects the collective excitations of the beads. The fact that the normalized experimental curves do not collapse to a single curve (Fig.~\ref{fig:dispersion_compare}a) reflects the change in pairing strength with $\kappa$, as was observed in the two-particle vortex (Fig.~\ref{fig:dtheta}). The sound velocity $C_s=(\partial\omega/\partial\kappa)_{k\rightarrow 0}$ approaches that of an equally distributed array of single particles at low pairing strength, and that of an equally distributed array of particle pairs at high pairing strength. It is interesting to note that the sound velocity of the particle waves ($180\mu$m/s$<C_s<810\mu$m/s) is higher than the average velocity of the particles ($12\mu$m/s$<v<150\mu$m/s). The shape of the experimental curve changes gradually with pairing strength approaching the single particle curve at weak pairing. For all cases, at low frequencies, or at large distances the dispersion relations are closer to those calculated for single particles, but at high frequencies, or short distances coupling prevails.
  
  Comparing our results with a driven 1D crystal oil droplet suspension studied recently \cite{beatus06,beatus07}, we find many similarities. Both systems exhibit symmetry breaking in the dispersion relations due to the directionality of driving. In both systems hydrodynamic interactions maintain vibrational modes in spite of the overdamped conditions, and the sound velocity of the collective vibrational modes is of the order of several hundreds of micrometers per second. 
  Unlike the oil droplet crystal, instabilities in the optical vortex are subdued due to the strong radial confinement. However, when the 1D volume fraction of particles, approaches $\phi_N\rightarrow0.5$ trapping becomes unstable. In such cases particles are continuously kicked out of the ring trap and are later on absorbed back in \cite{supp}. 
  Another unique trait of our system, is that hydrodynamic interactions are responsible both for creating the underlying configuration of equidistant particle pairs and for maintaining its vibrational excitations. This is different from the microfluidic 1D suspension whose crystalline structure stems from initial conditions of formation.
  Even in room temperature, where particle pairing is relatively weak (see Fig. \ref{fig:particles}a), we find that the pairing effect is dominant in the collective vibrations of the driven particles.
  
  The dynamics of the driven particles and their collective excitations are reproduced reasonably well by numerical simulation, as discussed above. However, the effect of thermal fluctuations on the dynamics should be examined in the future. While in simulation we expect thermal fluctuation to be decoupled from the periodic motion of the particles and hence result in a trivial cloaking of the motion, in experiments this need not be the case. The issue of whether thermal fluctuations or the vortex
  inhomogeneities are the dominant source of noise remains unclear and
  requires a more detailed study.
  
  The beating of large assemblies of cilia hairs can be random, synchronized (with  identical  phase), or metachronal (traveling wave) \cite{BLAKE1974,Tamm1975,niedermayer08}. Arrays of particles, each rotating around a ring, were used to model these collective modes both theoretically \cite{niedermayer08} and experimentally \cite{koumakis13}. Since particle pairing in a single vortex is related to the synchronized motion of two cilia hairs, it seems that the collective excitations observed in our system should correspond to collective beating of assemblies of cilia hairs. In this paper, the collective excitation found for many particles in a single ring is of the traveling wave mode, full synchronization would correspond to all particles moving in union in a single group. It would be interesting, to find the transition between the modes of motion in the optical vortex system and compare their onset to beating cilia, especially due to the larger number of particles accessible experimentally compared to multiple ring systems.      
  
\section*{Acknowledgments}

The authors wish to thank Roy Bar-Ziv, Haim Diamant, and Yair Shokef for fruitful discussions. In particular, we have benefited from extensive discussion with Tsevi Beatus. This research was supported by the James Franck German-Israeli Binational Program in Laser-Matter interactions, by the Israel Science Foundation grant 1271/08, and by the US-Israel Binational Science Foundation grant 2008483.


\end{document}